# On the Formulas for Quantum Mean Values for a Composite *A+B*

F. A R Navarro

farnape@gmail.com

Education National University, Lima 14 - Peru

**Abstract**

Herein is presented a research with regard to the calculation of quantum mean values, for a composite *A+B*, by using different formulas to expressions in Boltzmann-Gibbs-Shannon's statistics. It is analyzed why matrix formulas $E_A$ y $E_B$, in Hilbert subspaces, produce identical results to full Hilbert space formulas. In accord to former investigations, those matrices are the adequated density matrices, inside third version of nonextensive statistical mechanics. Those investigations were obtained by calculating the thermodynamical parameters of magnetization and internal energy. This publication demonstrates that it is not necessary postulate the mean values formulas in Hilbert subspaces, but they can be stem from full Hilbert space, taking into consideration the statistical independence concept.

*PACS*: 78.66.Sq, 05.30.-d, 98.38.Am, 75.10.Dg

**Keywords:** quantum statistical mechanics, composites, physical properties, density matrix, partial trace.

## I. INTRODUCTION

Nowadays, it is thought that nonextensive statistical mechanics could be an alternative for Boltzmann-Gibbs-Shannon's statistics. It was developed by Brazilian researcher C. Tsallis [1]. There are three versions about the quoted statistics; however some authors consider another fourth version [2]. In this article is analyzed the third version. So, herein is shown a generalization to the results got in *A study on Composed Nonextensive Magnetic Systems* [3], *New Consideration on Composed Nonextensive Magnetic Systems* [4] and *Internal Energy in the Context of Nonextensive Statistical Mechanics* [5]. Inside the preceding works, there was made an investigation about how must be the calculation of new formulas for quantum mean values, using Heisenberg model in mean field approach for a magnetic composite *A+B*. The density matrices $E_A$ and $E_B$ were consequence of former investigations. They resolved the calculations puzzle of mean value of one observable for different result happened if it was calculated in different Hilbert space.

In scientific literature, the nonextensive statistical mechanics has applications in many areas like Quantum Field Theory, Theory of Chaos, Condensate Matter Physics, Astrophysics, etc. [1].

## II. THEORETICAL FRAME

On 1988, C. Tsallis postulated the entropy $S_q$ [6]:

$$S_q = k_B \frac{1 - \text{Tr}(\hat{\rho}^q)}{q - 1}, \tag{1}$$

where $\hat{\rho}^q$ is the density operator $\hat{\rho}$ powered to entropic index $q$ (also called the Tsallis' entropic parameter), and $k_B$ is the Boltzmann's constant. **Tr** symbolizes the trace operation taken over the quantum states of the density matrix. When $q$ tends to 1 we recover the Boltzmann-Gibbs-Shannon's entropy:

$$S = k_B \text{Tr}(\hat{\rho} \text{Ln} \hat{\rho}), \tag{2}$$

The probability distribution $\rho$ is achieved through entropy maximization method which was proposed by E. T. Jaynes [7]. In that process should be utilized the following constraints of unitary norm and internal energy redefinition, $U_q$ [8]:

$$\text{Tr}(\hat{\rho}) = 1 \quad \text{and} \quad U_q = \frac{\text{Tr}(\hat{\rho}^q \hat{H})}{\text{Tr}(\hat{\rho}^q)}, \tag{3}$$

$\hat{H}$ stand for Hamiltonian operator. So it is obtained the probability density function (PDF):

$$\hat{\rho} = \frac{[\hat{1} - (1-q)\beta'\hat{H}]^{\frac{1}{1-q}}}{Z_q}, \tag{4}$$

where $Z_q$ is the generalized partition function,

$$Z_q = \text{Tr}[\hat{1} - (1-q)\beta'\hat{H}]^{\frac{1}{1-q}}, \tag{5}$$

and $\beta'$ is an energy parameter [9, 10].

On the other hand, for a composite $A+B$, we have the entropy of the complete system:

$$S_{A+B} = k_B \frac{1 - \text{Tr}_{(A,B)}[\hat{\rho}^q]}{q - 1}, \tag{6}$$

and for the subsystems $A$ and $B$:

$$S_A = k_B \frac{1 - \text{Tr}_A(\hat{\rho}_A^q)}{q - 1} \quad \text{and} \quad S_B = k_B \frac{1 - \text{Tr}_B(\hat{\rho}_B^q)}{q - 1}, \tag{7}$$

where **Tr$_A$** y **Tr$_B$** stand for, respectively, partial traces over states $A$ and $B$.

## 2.1 Quantum Mean Values

For a system $A+B$, in the third version of nonextensive statistical mechanics, the quantum mean values of observables $\hat{O}_A$ and $\hat{O}_B$, in the full Hilbert space, are calculated through formulas:

$$\frac{\text{Tr}_{(A,B)}(\hat{\rho}^q \hat{O}_A)}{\text{Tr}_{(A,B)}(\hat{\rho}^q)} \quad \text{and} \quad \frac{\text{Tr}_{(A,B)}(\hat{\rho}^q \hat{O}_B)}{\text{Tr}_{(A,B)}(\hat{\rho}^q)}, \tag{8}$$

and for the same quantum mean values, in Hilbert subspaces, the following expressions are taken into account:

$$\frac{\text{Tr}_A(\hat{\rho}_A^q \hat{\mathscr{O}}_A)}{\text{Tr}_A(\hat{\rho}_A^q)} \quad \text{and} \quad \frac{\text{Tr}_B(\hat{\rho}_B^q \hat{\mathscr{O}}_B)}{\text{Tr}_B(\hat{\rho}_B^q)}, \tag{9}$$

with the operators $\hat{\mathscr{O}}_A$ and $\hat{\mathscr{O}}_B$ representing, respectively, the subsystems observables $A$ and $B$, in Hilbert subspaces. But we also can use in place of last formulas yet new expressions, with the density matrices $E_A$ and $E_B$ [9, 10],

$$\hat{E}_A = \text{Tr}_B(\rho^q) \quad \text{and} \quad \hat{E}_B = \text{Tr}_A(\rho^q), \tag{10}$$

that is, now the quantum mean values are calculated in this way:

$$\frac{\text{Tr}_A(\hat{E}_A \hat{\mathscr{O}}_A)}{\text{Tr}_A(\hat{E}_A)} \quad \text{and} \quad \frac{\text{Tr}_B(\hat{E}_B \hat{\mathscr{O}}_B)}{\text{Tr}_B(\hat{E}_B)} \tag{11}$$

in previous publications it was clearly shown that matrices $E_A$ and $E_B$ are the adequated density matrices for calculations of quantum means of physical observables [4, 5].

## 2.2 Statistical Independence

In the Boltzmann-Gibbs-Shannon's statistical mechanics, two systems, $A$ and $B$, are statistically independent if the following relation is verified:

$$\hat{\rho}_{(A,B)} = \hat{\rho}_A \otimes \hat{\rho}_B, \tag{12}$$

where the symbol $\otimes$ is the tensor product operator.

Extending this expression to nonextensive statistical mechanics, we have:

$$\hat{\rho}^q = \hat{\rho}_A^q \otimes \hat{\rho}_B^q, \tag{13}$$

which can also be expressed so:

$$\text{Tr}_{(A,B)}(\hat{\rho}^q) = \text{Tr}_A[\hat{\rho}_A^q]\text{Tr}_B[\hat{\rho}_B^q], \tag{14}$$

this formulation will be outstanding to understand the procedure that we display next.

**III. PROCEDURE**

Like it was mentioned, in the references [3, 4, 5] were exposed exact analytical calculations of quantum mean values regarding the suitableness of using the matrices $E_A$ and $E_B$. However, these publications display how operate with matrices $E_A$ and $E_B$. There was lost an explanation about why these matrices produce identical calculation so much full Hilbert space so Hilbert subspaces. Then, we next demonstrate how by passing from observables formulas in full Hilbert space to formulas in Hilbert subspaces. A more explicit calculation for $q=1$ is found in the classical book of quantum mechanics [11]. For didactic reasons we have simplified the calculations.

So in the following scheme, we have four steps to going from formulas in full Hilbert space to formulas in Hilbert subspaces. Despite its apparent simplicity, it shows to be a powerful tool for our task.

$$
\begin{array}{cccc}
1 & 2 & 3 & 4 \\
\end{array}
$$

$$\rho^q = \rho_A^q \otimes \rho_B^q$$
$$\downarrow$$

$$\frac{\text{Tr}_{(A,B)}(\rho^q \hat{O}_A)}{\text{Tr}_{(A,B)}(\rho^q)} \Rightarrow \frac{\text{Tr}_A(\hat{E}_A \hat{\mathcal{O}}_A)}{\text{Tr}_A(\hat{E}_A)} \Rightarrow \frac{\text{Tr}_B(\rho_B^q)\text{Tr}_A(\rho_A^q \hat{\mathcal{O}}_A)}{\text{Tr}_B(\rho_B^q)\text{Tr}_A(\rho_A^q)} \Rightarrow \frac{\text{Tr}_A(\rho_A^q \hat{\mathcal{O}}_A)}{\text{Tr}_A(\rho_A^q)} \text{ and}$$

$$\frac{\text{Tr}_{(A,B)}(\rho^q \hat{O}_B)}{\text{Tr}_{(A,B)}(\rho^q)} \Rightarrow \frac{\text{Tr}_B(\hat{E}_B \hat{\mathcal{O}}_B)}{\text{Tr}_B(\hat{E}_B)} \Rightarrow \frac{\text{Tr}_A(\rho_A^q)\text{Tr}_B(\rho_B^q \hat{\mathcal{O}}_B)}{\text{Tr}_A(\rho_A^q)\text{Tr}_B(\rho_B^q)} \Rightarrow \frac{\text{Tr}_B(\rho_B^q \hat{\mathcal{O}}_B)}{\text{Tr}_B(\rho_B^q)}$$

In this passage from the formulas in Full Hilbert space to the formulas in Hilbert subspaces, we have: The step **1** represents the quantum mean value of an observable, respectively, in subsystem *A* and *B*, in full Hilbert space, all operator are in this space. The step **2** is obtained when we carrying out the respective partial traces over states *B* and *A*, that is, the following operators are got through partial traces: $\hat{E}_A = \text{Tr}_B(\hat{\rho}^q)$ and $\hat{E}_B = \text{Tr}_A(\hat{\rho}^q)$, also $\hat{\mathcal{O}}_A = \text{Tr}(\hat{O}_A)$ and $\hat{\mathcal{O}}_B = \text{Tr}_A(\hat{O}_B)$, these operators are represented in the respective Hilbert subspaces. The step **3** is obtained from the step **2** if we assume statistical independence, that is, $\hat{\rho}^q = \hat{\rho}_A^q \otimes \hat{\rho}_B^q$. And, finally, the step **4** contains the simplified formulas in Hilbert subspaces, these are the usual postulated formulas.

## IV. CONCLUSIONS

In this paper, initially, in the frame of third version of nonextensive statistical mechanics, utilizing the partial traces concept, there were demonstrated that the mean values formulas for a composite *A+B*, in Hilbert subspaces, can be deduced (but not necessarily postulated) from the formulas in the full Hilbert space. For this procedure is very important the hypothesis of statistical independence between the subsystems *A* and *B*. So, this article generalizes the particular results obtained in the references [3, 4, 5]. Additionally, it naturally is concluded that the passage from formulas in full Hilbert space to Hilbert subspaces is independent of temperature, entropic index, density matrix, and it stays valid for any physical observable. Finally, these results are can be generalized beyond nonextensive statistical mechanics.

## V. BIBLIOGRAFY